\documentclass[useAMS,usenatbib,usegraphicx]{mn2e}

\newcommand{\A}[1]{\bmath{#1}}
\newcommand{\sign}{\textrm{sign}}
\newcommand{\pd}[2]{\frac{\partial #1}{\partial #2}}
\newcommand{\DS}{\displaystyle}

\newcommand{\HALF}{\frac{1}{2}}

\newtheorem{theorem}{Theorem}[section]

\newtheorem{proposition}[theorem]{Proposition}

\newenvironment{demonstration}[1][Proof]{\begin{trivlist}
\item[\hskip \labelsep {\bfseries #1}]}{\end{trivlist}}

\newcommand{\qed}{\nobreak \ifvmode \relax \else
      \ifdim\lastskip<1.5em \hskip-\lastskip
      \hskip1.5em plus0em minus0.5em \fi \nobreak
      \vrule height0.75em width0.5em depth0.25em\fi}

\title[An HLLC Riemann Solver for Relativistic Flows: I. Hydrodynamics]
      {An HLLC Riemann Solver for Relativistic Flows: I. Hydrodynamics}
\author[A. Mignone and G. Bodo]{A. Mignone$^{1}$\thanks{E-mail:
mignone@to.astro.it}        and G. Bodo$^{1}$\\
$^{1}$INAF Osservatorio Astronomico di Torino, 10025 Pino Torinese}
\begin{document}

\date{Accepted ??. Received ??; in original form ??}

\pagerange{\pageref{firstpage}--\pageref{lastpage}} \pubyear{2005}

\maketitle

\label{firstpage}

\begin{abstract}
  We present an extension of the HLLC approximate Riemann solver
  by Toro, Spruce and Speares  
  to the relativistic equations of fluid dynamics.
  The solver retains the simplicity of the original two-wave 
  formulation proposed by Harten, Lax and van Leer (HLL) but 
  it restores the missing contact wave in the solution of 
  the Riemann problem.
  The resulting numerical scheme is computationally efficient, 
  robust and positively conservative. 
  The performance of the new solver is evaluated through
  numerical testing in one and two dimensions. 
\end{abstract}


%%%%%%%%%%%%%%%%%%%%%%%%%%%%%%%%%%%%%%%%%%%%%%%%%%%%%%%%%%%%%%%%%%%%%%%%%%%%%
\section{Introduction}
%
%
%
%
%
%%%%%%%%%%%%%%%%%%%%%%%%%%%%%%%%%%%%%%%%%%%%%%%%%%%%%%%%%%%%%%%%%%%%%%%%%%%%%

High energy astrophysical phenomena involve, in many cases,
relativistic flows, typical examples are superluminal motion of
relativistic jets in extragalactic radio sources, accretion flows
around massive compact objects, pulsar winds and Gamma Ray Bursts. The
modeling of such phenomena has prompted the search for efficient and
accurate numerical formulations of the special relativistic fluid
equations \citep[for an excellent review see][]{MM03}. There is now
a strong consensus that the so-called ``high-resolution
shock-capturing'' schemes 
provide the necessary tools in developing stable
and robust relativistic fluid dynamical codes. 
One of the fundamental ingredients of such schemes is the 
exact or approximate solution to the Riemann problem. 

The solution to the Riemann problem in relativistic 
hydrodynamics (RHD henceforth) has been extensively 
studied in literature, and an exact solution can be 
found within high degree of accuracy by iterative techniques, see 
\citet{MM94}, \citet{PMM00}, \citet{RZP03} and references therein.
One of the major differences with the classical counterpart
is the velocity coupling introduced by the Lorentz factor 
and the coupling of the latter with the specific enthalpy.
This considerably adds to the computational cost, 
making the use of an exact solver code prohibitive in 
a multidimensional Godunov-type code.

From this perspective, approximate solvers based  
on alternative strategies have been devised:
local linearization \citep{EM95, FK96}, two-shock approximation
\citep{Balsara94, DW97, MPB05}, flux-splitting methods
\citep{DFIM98}, and so forth, see \citet{MM03} for a 
comprehensive review. 
Most of these solvers, however, rely on rather expensive 
characteristic decompositions of the Jacobian matrix 
or involve iterative techniques to solve highly nonlinear
equations.
Although they usually attain better resolution at discontinuities,
some of these methods may produce unphysical states with negative
densities or pressures, as it has been shown by \citet{EMRS91}
for linearized Riemann solvers in the context of classical 
hydrodynamics.

The HLL method devised by \citet{HLL83} for classical gasdynamics 
belongs to a different class of approximate Riemann solvers and 
has gained increasing popularity among researchers in the last decade.
It has been implemented in the context of the relativistic fluid
equations by \citet{SKRWMM93} and \citet{DH94}.
The HLL approach does not require a full characteristic 
decomposition of the equations and is straightforward to 
implement in a Godunov-type code.
Besides the computational efficiency, this class of solvers has 
the attractive feature of being positively conservative
in the sense that preserve initially positive densities, energy 
and pressures.   

The HLL formulation, however, lacks the ability to resolve 
isolated contact discontinuity and for this reason has 
a more diffusive character than other more sophisticated algorithms.
To compensate for this, \citet{TSS94} developed an extension of the HLL 
solver for the Euler equations introducing a two-state HLL-type 
solver called HLLC (where ``C'' stands for contact) that improves
the treatment of the contact discontinuity, see also \citet{BCLC97}.
Recently this approach has been generalized to the magnetohydrodynamic
equations \citep{Gurski04,Li05}.

In the present work, we extend this approach to the relativistic
equation of fluid dynamics. 
The paper is structured as follows: in \S\ref{sec:equations} 
the relevant equations are given, in \S\ref{sec:hll} we
describe the new approximate Riemann solver and in \S\ref{sec:test}
we asses the strength of the new method with one and two dimensional 
tests.
    
%%%%%%%%%%%%%%%%%%%%%%%%%%%%%%%%%%%%%%%%%%%%%%%%%%%%%%%%%%%%%%%%%%%%%%%%%%%%%
\section{The RHD Equations}\label{sec:equations}
%
%
%
%
%
%%%%%%%%%%%%%%%%%%%%%%%%%%%%%%%%%%%%%%%%%%%%%%%%%%%%%%%%%%%%%%%%%%%%%%%%%%%%%

The motion of an ideal relativistic fluid is governed by
conservation of mass, momentum and energy. 
The pertaining equations are cast as a hyperbolic system of 
conservation laws \citep{Lan_Lif59} which, in two dimensions, reads
\begin{equation}\label{eq:rhd_eq}
 \pd{\A{U}}{t} + \pd{\A{F}^x(\A{U})}{x} + \pd{\A{F}^y(\A{U})}{y} = 0 \, ,
\end{equation}
where $\A{U} = (D, m_x, m_y, E)$ is the unknown vector of conservative 
variables, whereas $\A{F}^x$ and $\A{F}^y$ are, respectively,
the fluxes along the $x$ and $y$ directions:
\begin{equation}\label{eq:vectors}
 \A{F}^x(\A{U}) = \left(\begin{array}{c}
             Dv_x    \\ \noalign{\medskip}
             m_x v_x + p \\ \noalign{\medskip}
             m_y v_x     \\ \noalign{\medskip}
             m_x   \\ \noalign{\medskip}
        \end{array} \right) \, , \quad
 \A{F}^y(\A{U}) = \left(\begin{array}{c}
             Dv_y    \\ \noalign{\medskip}
             m_x v_y \\ \noalign{\medskip}
             m_y v_y + p \\ \noalign{\medskip}
             m_y   \\ \noalign{\medskip}
        \end{array} \right) \, .
\end{equation}
Generalization to three dimensions is straightforward.

In equations (\ref{eq:vectors}), $p$ is the thermal 
pressure, whereas $D$, $\A{m}\equiv (m_x, m_y)$ and $E$ 
are, respectively, the mass, momentum and energy densities 
relative to the lab frame, where the fluid has velocity
$\A{v} \equiv (v_x, v_y)$.
Units are conveniently normalized so that the speed 
of light is $c = 1$.

The relation between conserved variables $\A{U}$ and 
physical quantities $\A{V} = \big(\rho, v_x, v_y, p\big)$ is
\begin{equation}\label{eq:p2c}
    D   = \gamma \rho \,, \quad  \A{m} = Dh\gamma\A{v} \,,
    \quad E   = Dh\gamma - p\,, 
\end{equation}
where $\rho$ is the proper rest mass density, 
$\gamma = (1 - \A{v}\cdot\A{v})^{-\HALF}$ is the Lorentz
factor and $h$ is the specific enthalpy.
Proper closure is provided by specifying an equation of state
in the form $h = h(p, \rho)$. 

For an ideal gas, the enthalpy has the form 
$\rho h = \rho + p\Gamma/(\Gamma - 1)$ and 
the sound speed is defined by
\begin{equation}\label{eq:sound_speed}
  c_s = \sqrt{\frac{\Gamma p}{\rho h}} \,.
\end{equation}
with $\Gamma$ being the (constant) specific heat ratio.
By letting $p/\rho\to\infty$, we see that the square of the sound speed
has the limiting value $c_s^2 \to \Gamma - 1$.
Since it can be shown \citep{Taub48,Anile89,MPB05} that the specific 
heat ratio $\Gamma$ cannot exceed $2$, one always has $c_s^2 < 1$. 
This is an important result for the positivity of the HLL and HLLC schemes 
and will be used in a later section.

Equation (\ref{eq:p2c}) gives $\A{U}$ in terms of the 
primitive state vector $\A{V}$. The inverse relation
involves the solution of a nonlinear equation for the pressure $p$:
\begin{equation}\label{eq:c2p}
 E + p = D\gamma + \frac{\Gamma}{\Gamma - 1}p\gamma^2
\end{equation}  
where $\gamma = \big[1 - |\A{m}|^2/(E + p)^2\big]^{-\HALF}$.
Equation (\ref{eq:c2p}) can be solved by any standard root finding 
algorithm.

%%%%%%%%%%%%%%%%%%%%%%%%%%%%%%%%%%%%%%%%%%%%%%%%%%%%%%%%%%%
\subsection{The Riemann Problem in RHD}
%
%
%%%%%%%%%%%%%%%%%%%%%%%%%%%%%%%%%%%%%%%%%%%%%%%%%%%%%%%%%%%

Consider a conservative discretization of (\ref{eq:rhd_eq}) 
along the x-direction:
\begin{equation}\label{eq:update}
 \frac{\bar{\A{U}}^{n+1}_i - \bar{\A{U}}^{n}_i}{\Delta t^n} =
 \frac{\A{f}_{i+\HALF} - \A{f}_{i-\HALF}}{\Delta x_i}  \,.
\end{equation} 

The numerical flux functions $\A{f}_{i+\HALF}$ follow from
the solution of Riemann problems with initial data:
\begin{equation}\label{eq:riemann}
 \A{U}(x,0) = \left\{\begin{array}{ccc}
   \A{U}_{L,i+\HALF} & \quad \textrm{if} \; & x < x_{i+\HALF} \,, \\ \noalign{\medskip}
   \A{U}_{R,i+\HALF} & \quad \textrm{if} \; & x > x_{i+\HALF} \,, \\ \noalign{\medskip}
\end{array}\right.
\end{equation}
where $\A{U}_{L,i+\HALF}$ and $\A{U}_{R,i+\HALF}$ are the left
and right edge values at zone interfaces.

\begin{figure}
 \includegraphics[width=84mm]{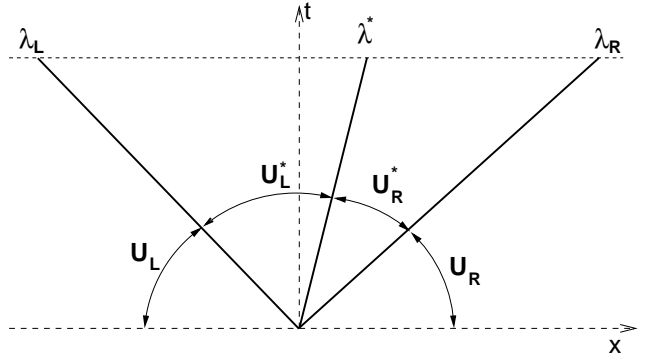}
 \caption{Graphical representation of the Riemann fan in the $x-t$ plane. 
          The two initial states $\A{U}_L$ and $\A{U}_R$ decay 
          into two nonlinear waves (with speeds $\lambda_L$ and 
          $\lambda_R$) and a linear contact wave with velocity 
          $\lambda^*$. The resulting wave pattern divides the 
          $x-t$ plane into four regions each defining
          a constant state: $\A{U}_L$, $\A{U}^*_L$, $\A{U}^*_R$ and $\A{U}_R$.}
 \label{fig:fan}
\end{figure}

The solution of the Riemann problem for the special relativistic 
fluid equations has been investigated by \citet{MM94}, \citet{PMM00}, 
\citet{RZP03}.
It consists of a self-similar three-wave pattern 
generated by the decay of the initial discontinuity (\ref{eq:riemann}).
The resulting Riemann fan (Fig. \ref{fig:fan}) is bounded by two nonlinear waves
(representing either shocks or rarefactions) separated by a contact
discontinuity moving at the fluid velocity.
Across the contact discontinuity, pressure and normal velocity are
continuous whereas density and tangential velocities experience
jumps. 
The same holds also in the non-relativistic limit.
Contrary to the Newtonian counterpart, however, 
all variables are discontinuous across a shock wave or change 
smoothly through a rarefaction fan \citep{PMM00}.
This is a consequence of the velocity coupling introduced by 
the Lorentz factor $\gamma$ and by the coupling of the latter with 
the specific enthalpy $h$.

The resulting wave pattern can be solved to a high degree of
precision by iterative techniques and has been implemented 
for the first time in the one-dimensional Godunov-type code 
by \citet{MM96}.
Nevertheless, when tangential velocities are included, the computational
effort increases considerably, and the use of an exact solver 
in a multidimensional Godunov-type code can become prohibitive. 

Here we consider a different approach, based on the original
prescription given by \citet{HLL83} (HLL) for the classical Euler
equations and subsequently 
extended by \citet{TSS94} (HLLC).
Both the HLL and HLLC formulations do not require 
a field by field decomposition of the relativistic equations,
a feature which makes them particularly attractive, specially in
multi-dimensional applications.

%%%%%%%%%%%%%%%%%%%%%%%%%%%%%%%%%%%%%%%%%%%%%%%%%%%%%%%%%%%%%%%%%
\section{The HLL Framework}\label{sec:hll}
%
%
%%%%%%%%%%%%%%%%%%%%%%%%%%%%%%%%%%%%%%%%%%%%%%%%%%%%%%%%%%%%%%%%

Harten, Lax and van Leer \citep{HLL83} proposed an approximate 
solution to the Riemann problem where the two states bounded by the two 
acoustic waves are averaged into a single constant state. 
In other words, the solution to the Riemann problem on the $x/t = 0$ axis 
consists of the three possible constant states:
\begin{equation}\label{eq:hll_states}
  \A{U}(0,t) = \left\{\begin{array}{cll}
     \A{U}_L      & \; \textrm{if} \; & \lambda_L \ge 0 \,, \\ \noalign{\medskip}
     \A{U}^{hll}  & \; \textrm{if} \; & \lambda_L \le 0 \le \lambda_R \,, \\ \noalign{\medskip}
     \A{U}_R      & \; \textrm{if} \; & \lambda_R \le 0 \,, \\ \noalign{\medskip}
\end{array}\right.
\end{equation}
where we dropped, for simplicity, the half integer notation $i+\HALF$. 
\citet{HLL83} noted that the single state $\A{U}^{hll}$ could be 
constructed from an a priori estimate of the fastest and slowest signal 
velocities $\lambda_L$ and $\lambda_R$:
\begin{equation}\label{eq:hll_state}
 \A{U}^{hll} = \frac{\lambda_R \A{U}_R - \lambda_L\A{U}_L + 
                      \A{F}_L - \A{F}_R}{\lambda_R - \lambda_L} \,,
\end{equation}
where $\A{F}_L = \A{F}^x(\A{U}_L)$, $\A{F}_R = \A{F}^x(\A{U}_R)$.
Notice that equation (\ref{eq:hll_state}) represents the 
integral average of the solution of the Riemann problem
over the wave fan \citep{Toro97}.

The corresponding interface numerical flux is defined as:
\begin{equation}\label{eq:hll_fluxes}
 \A{f} = \left\{\begin{array}{cll}
   \A{F}_L     & \;  \textrm{if} \;&   \lambda_L \ge 0 \,, \\ \noalign{\medskip}
   \A{F}^{hll} & \;  \textrm{if} \;&   \lambda_L \le 0 \le \lambda_R \,, \\ \noalign{\medskip}
   \A{F}_R     & \;  \textrm{if} \;&   \lambda_R \le 0 \,, \\ \noalign{\medskip}
\end{array}\right.
\end{equation}
where
\begin{equation}\label{eq:hll_flux}
 \A{F}^{hll} = \frac{\lambda_R\A{F}_L - \lambda_L\A{F}_R + \lambda_R\lambda_L (\A{U}_R - \A{U}_L)}
            {\lambda_R - \lambda_L}   \,.
\end{equation}

Thus, given a wave speed estimate for the fastest and slowest speeds 
$\lambda_R$ and $\lambda_L$ (see \S\ref{sec:speeds}), an 
approximate solution to the Riemann problem can be constructed 
and the intercell numerical fluxes for the conservative update 
(\ref{eq:update}) are computed using (\ref{eq:hll_fluxes}).

This approach has been applied for the first time to the 
one-dimensional relativistic equations by \citet{SKRWMM93}
and later by \citet{DH94} for the multidimensional case.

Although the HLL prescription is computationally inexpensive and 
straightforward to implement, a major drawback is its inability 
to resolve contact or tangential waves.
On the contrary the HLLC scheme, originally introduced 
by \citet{TSS94} in the context of the Euler equations of classical 
gasdynamics, does not suffer from this loss. 
In the next section we generalize this approach to the equations
of relativistic hydrodynamics.
 
%%%%%%%%%%%%%%%%%%%%%%%%%%%%%%%%%%%%%%%%%%%%%%%%%%%%%%%%%%%%%%%%%%%%%%%%%%%%%
\subsection{HLLC Solver}\label{sec:hllc}
%
%
%
%
%
%%%%%%%%%%%%%%%%%%%%%%%%%%%%%%%%%%%%%%%%%%%%%%%%%%%%%%%%%%%%%%%%%%%%%%%%%%%%%

The HLLC scheme restores the full wave structure inside the Riemann fan
by replacing the single averaged state defined by (\ref{eq:hll_state}) with 
two approximate states, $\A{U}^*_L$ and $\A{U}^*_R$.
These two states are separated by a middle contact wave which is assumed 
to have constant speed $\lambda^*$, so that the full solution
to the Riemann problem now reads 
\begin{equation}\label{eq:hllc_states}
  \A{U}(0,t) = \left\{\begin{array}{ccc}
   \A{U}_L    & \quad \textrm{if} & \; \lambda_L \ge 0    \,,              \\ \noalign{\medskip}
   \A{U}^*_L  & \quad \textrm{if} & \; \lambda_L \le 0 \le \lambda^* \,,\\ \noalign{\medskip}
   \A{U}^*_R  & \quad \textrm{if} & \; \lambda^* \le 0 \le \lambda_R \,,\\ \noalign{\medskip}
   \A{U}_R    & \quad \textrm{if} & \; \lambda_R \le 0              \,, \\ \noalign{\medskip}
\end{array}\right. 
\end{equation}
and the corresponding intercell numerical fluxes are: 
\begin{equation}\label{eq:hllc_flux}
 \A{f} = \left\{\begin{array}{ccc}
   \A{F}_L    & \quad \textrm{if} & \; \lambda_L \ge 0              \,, \\ \noalign{\medskip}
   \A{F}^*_L  & \quad \textrm{if} & \; \lambda_L \le 0 \le \lambda^*\,, \\ \noalign{\medskip}
   \A{F}^*_R  & \quad \textrm{if} & \; \lambda^* \le 0 \le \lambda_R\,, \\ \noalign{\medskip}
   \A{F}_R    & \quad \textrm{if} & \; \lambda_R \le 0   \,.          \\ \noalign{\medskip}
\end{array}\right. 
\end{equation}

The intermediate state fluxes $\A{F}^*_L$ and $\A{F}^*_R$ may be 
expressed in terms of $\A{U}^*_L$ and $\A{U}^*_R$ from the 
Rankine-Hugoniot jump conditions:
\begin{equation}\label{eq:jump_1}
  \lambda \left(\A{U}^* - \A{U}\right) = 
  \A{F}^* - \A{F} \,,
\end{equation}
where here and throughout the following, quantities without a suffix $L$ or 
$R$ refer indifferently to the left ($L$) or right ($R$) states.
Notice that, in general, $\A{F} = \A{F}^x(\A{U})$ but 
$\A{F}^* \neq \A{F}^x(\A{U}^*)$.

We remind the reader that the HLL and HLLC solvers differ in the 
representation of the intermediate states. In the case of 
supersonic flows ($\lambda_L > 0$ or $\lambda_R < 0$), in fact,
the two solvers become equivalent. The same result also holds for 
an exact Riemann solver.

If $\lambda_L$ and $\lambda_R$ are given (see \S\ref{sec:speeds}), 
equation (\ref{eq:jump_1}) represent a system of $2n$ equations 
(where $n$ is the number of components of $\A{U}$) for the 
$4n + 1$ unknowns $\A{U}^*_L$,  $\A{U}^*_R$, $\A{F}^*_L$, $\A{F}^*_R$ 
and $\lambda^*$.
Three additional constraints come from the requirements that 
both pressure and normal velocity be continuous across the 
contact wave (i.e. $v^*_{x,R} = v^*_{x,L}$, 
$p^*_R = p^*_L$) and that  
$\lambda^* = v_{x,L}^* = v_{x,R}^*$. This, however, 
yields a total of only $2n + 3$ equations, still not sufficient
to solve the system. 
In order to reduce the number of unknowns and have a 
well-posed problem, further assumptions
have to be made on the form of the fluxes $\A{F}^*$.
Here we assume that the two-dimensional fluxes can be 
written as
\begin{equation}\label{eq:flux}
 \A{F}^* = \left(\begin{array}{c}
   D^*v_x^*         \\ \noalign{\medskip}
   m_x^*v_x^* + p^*  \\ \noalign{\medskip}
   m_y^*v_x^*        \\ \noalign{\medskip}
   m_x^*           
\end{array}\right) \,.
\end{equation}

In such a way, both $\A{U}^*$ and $\A{F}^*$ are expressed in terms of 
the five unknowns $D^*$, $v_x^*$, $m_y^*$, $E^*$ and $p^*$. 
The normal components of momentum in the star region, $m^*_{x,L}$
and $m^*_{x,R}$, 
are not independent variables since, for consistency, we require 
that $m_x^* = (E^* + p^*)v_x^*$. In the classical case,
this assumption becomes equivalent to $m^*_x = \rho^*\lambda^*$.
This yields a total of $11$ equations in $11$ unknowns. 

Writing explicitly equation (\ref{eq:jump_1}) for the left
or the right state yields
\begin{equation}\label{eq:jumps}
 \begin{array}{lcl}
 D^* (\lambda - \lambda^*)  & = & D(\lambda - v_x)             \,, \\ \noalign{\medskip}
 m_x^*(\lambda - \lambda^*) & = & m_x(\lambda - v_x) + p^* - p \,, \\ \noalign{\medskip}
 m_y^*(\lambda - \lambda^*) & = & m_y(\lambda - v_x)  \,, \\ \noalign{\medskip}
 E^*(\lambda - \lambda^*)   & = & E(\lambda - v_x) + p^*\lambda^* - pv_x \,.
\end{array}
\end{equation}

If one combines the last of (\ref{eq:jumps}) together with the 
second one, the following expression giving $\lambda^*$ in terms of 
$p^*$ may be obtained:
\begin{equation}\label{eq:lambda_p}
 (A - \lambda p^*)v_x^* = B + p^*  \,,
\end{equation}
where $A = \lambda E - m_x$, $B = m_x(\lambda - v_x) - p$.

By imposing $p^*_{x,L} = p^*_{x,R}$ across the contact 
discontinuity we find the following quadratic equation for $\lambda^*$:
\begin{equation}\label{eq:lambdastar}
   F_E^{hll} \left(\lambda^*\right)^2 
 - (E^{hll} + F_{m_x}^{hll})\lambda^* 
 + m_x^{hll} = 0 \,.
\end{equation}

In equation (\ref{eq:lambdastar}), $F_E^{hll}$ and $F_{m_x}^{hll}$ 
are the energy and momentum
components of the HLL flux given by equation (\ref{eq:hll_flux}),
whereas $E^{hll}$ and $m_x^{hll}$ are the energy and 
normal momentum components of the HLL state vector, 
equation (\ref{eq:hll_state}).
Of the two roots of equation (\ref{eq:lambdastar}) only the one
with the minus sign is physically acceptable, since it lies in the 
range $(-1,1)$ and, according to the wave-speed estimate presented 
in \S\ref{sec:speeds}, can be interpreted as an
average velocity between $\lambda_L$ and
$\lambda_R$. The mathematical proof of this statement is given in 
the appendix (\ref{app:A}).
Besides, the same root has the correct classical limit, that is 
$\lambda^* \to m_x^{hll}/\rho^{hll}$ as $v/c\to 0$ and $h\to 1$.
This wave speed is the same one proposed by \citet{Toro97} and
further discussed in \citet{BCLC97}.

Once $\lambda^*$ is known, $p^*$ is computed from (\ref{eq:lambda_p}) 
and the components of $\A{U}^*$ are readily obtained from (\ref{eq:jumps}).
 
Finally we notice that the method is consistent, in that  
the integral average over the Riemann fan automatically satisfies the 
consistency condition by construction \citep{Toro97}:
\begin{equation}\label{eq:consistency}
 \frac{(\lambda^* - \lambda_L) \A{U}^*_L + 
 (\lambda_R - \lambda^*) \A{U}^*_R}{\lambda_R - \lambda_L} = \A{U}^{hll}  \,,
\end{equation}  
or, alternatively,
\begin{equation}\label{eq:consistency2}
 \frac{\A{F}^*_L\lambda_R(\lambda^* - \lambda_L) + 
   \A{F}^*_R\lambda_L(\lambda_R - \lambda^*)}{\lambda_R - \lambda_L} = \lambda^*\A{F}^{hll}\,.
\end{equation}
Incidentally, we notice that equation (\ref{eq:lambdastar}) could have 
been obtained by algebraic manipulations of 
equations (\ref{eq:consistency}) and (\ref{eq:consistency2}).

%%%%%%%%%%%%%%%%%%%%%%%%%%%%%%%%%%%%%%%%%%%%%%%%%%%%%%%%%%%%%%
\subsubsection{Wave Speed Estimate}\label{sec:speeds}
%
%
%
%%%%%%%%%%%%%%%%%%%%%%%%%%%%%%%%%%%%%%%%%%%%%%%%%%%%%%%%%%%%%%

The wave speeds needed in our formulation are estimates for
the lower and upper bounds of the signal velocities in the solution
to the Riemann problem \citep{Toro97}.
Here we consider the relativistic generalization of the estimates
given by \citet{Davis88} for the Euler equation of gasdynamics.  
The same choice has been initially adopted by \citet{SKRWMM93}, \citet{DH94} 
in their relativistic HLL solver and is commonly used by other authors, see,
for example, \citet{dZB02}.
Specifically we set:
\begin{equation}\label{eq:wavespeeds}
 \begin{array}{c}
 \lambda_L = \min(\lambda_-(\A{V}_R), \lambda_-(\A{V}_L))  \,, 
 \\ \noalign{\medskip}
  \lambda_R = \max(\lambda_+(\A{V}_R), \lambda_+(\A{V}_L))  \,,
 \end{array}
\end{equation}
where $\lambda_+$ and $\lambda_-$ are the maximum and minimum 
eigenvalues of the Jacobian matrix $\partial\A{F}/\partial\A{U}$.
They are the roots of the quadratic equation
\begin{equation}\label{eq:eigenvalues_eq}
 (\lambda - v_x)^2 = \sigma_s(1 - \lambda^2) \,,
\end{equation}
with $\sigma_s = c_s^2/\big(\gamma^2(1-c_s^2)\big)$,
and hence
\begin{equation}\label{eq:eigenvalues}
  \lambda_\pm(\A{V}) = \frac{v_x \pm \sqrt{\sigma_s\left(1 - v_x^2 + \sigma_s\right)}}{1+\sigma_s}\,.
\end{equation}

It should be mentioned that the wave speed estimate (\ref{eq:wavespeeds})
is not the only possible one and different choices (such as the Roe average) 
may be considered. 

%%%%%%%%%%%%%%%%%%%%%%%%%%%%%%%%%%%%%%%%%%%%%%%%%%%%%%%%%%%%%%%%%%%%%
\subsubsection{Positivity of the HLLC scheme}\label{sec:positivity}
%
%
%
%
%%%%%%%%%%%%%%%%%%%%%%%%%%%%%%%%%%%%%%%%%%%%%%%%%%%%%%%%%%%%%%%%%%%%%

Adopting the same notations as in \citet{BCLC97}, we denote with $G$
the set of physically admissible conservative states:
\begin{equation}\label{eq:positivity}
 G = \left\{ \left(\begin{array}{c}
               D   \\  \noalign{\medskip}
               m_x \\  \noalign{\medskip}
               m_y \\  \noalign{\medskip}
               E   \end{array}\right) \,,\quad
           \begin{array}{l}   
            D > 0  \\ \noalign{\bigskip}
            E > \sqrt{m_x^2 + m_y^2 + D^2} 
           \end{array}
 \right\} \,,
\end{equation}
where the second inequality simultaneously guarantees 
pressure positivity and that the total velocity never
exceeds the speed of light.

We remind the reader that the pressure $p(\A{U}^*)$ 
computed from the conservative state $\A{U}^*$ using 
(\ref{eq:c2p}) should not be confused with $p^*$ appearing   
in the flux definition (\ref{eq:flux}). 
The two pressures are, in fact, different and the 
positivity argument should apply to $p(\A{U}^*)$
rather than $p^*$, which can take negative values
under certain circumstances.
Similar considerations hold for the velocity $\lambda^*$ of 
the contact discontinuity for which, in general, we have 
$\lambda^* \neq v_x(\A{U}^*)$.
Thus $p^*$ and $\lambda^*$ may be more conveniently considered as
auxiliary variables. 

This is one of the fundamental differences between our relativistic
solver and the classical HLLC scheme, for which $\lambda^* = m_x^{hll}/\rho^{hll}$
and thus only $p^*$ plays the role of an auxiliary parameter.
This behavior is a direct consequence of the 
relativistic coupling between thermodynamical and kinetical terms, 
a feature absent in the Newtonian formulation.
              
The positivity of the HLLC scheme is preserved if each 
of the two intermediate states $\A{U}^*_L$ and $\A{U}^*_R$ are contained 
in G.  

For the density, the proof is trivial and follows from the
inequalities $\lambda_L\le\lambda^*\le\lambda_R$ 
and $\lambda_L\le v_x (L,R) \le \lambda_R$, see Appendix \S\ref{app:A}.

Unfortunately the analytical proof of the second statement presents
some algebraic difficulties, since the second of (\ref{eq:positivity})
reduces to an inequality for a quartic equation in $\lambda^*$. 
However, extensive numerical testing, part of which is presented in 
\S\ref{sec:test}, has shown that the second of (\ref{eq:positivity}) is
always satisfied for all pair of states $\A{U}_L$ and $\A{U}_R$ whose 
wave speeds are computed according to (\ref{eq:wavespeeds}) 
and for which an exact analytical solution to the Riemann 
problem exists (i.e. no vacuum is created).

%%%%%%%%%%%%%%%%%%%%%%%%%%%%%%%%%%%%%%%%%%%%%%%%%%%%%%%%%%%%%%%%%%%%%
\section{Algorithm Validation}\label{sec:test}
%
%
%
%
%%%%%%%%%%%%%%%%%%%%%%%%%%%%%%%%%%%%%%%%%%%%%%%%%%%%%%%%%%%%%%%%%%%%%

We now provide some numerical examples to test our new HLLC solver.
For the test problems considered in this section we closely 
follow \citet{SFIM04}.
 
%%%%%%%%%%%%%%%%%%%%%%%%%%%%%%%%%%%%%%%%%%%%%%%%%%%%%%%%%%%%%%%%
\subsection{Implementation Details}
%
%
%%%%%%%%%%%%%%%%%%%%%%%%%%%%%%%%%%%%%%%%%%%%%%%%%%%%%%%%%%%%%%%%

The numerical integration of the relativistic equations 
(\ref{eq:rhd_eq}) proceeds via the conservative update (\ref{eq:update}).
For the first-order HLLC scheme, we compute the inter-cell numerical 
fluxes $\A{f}_{i+\HALF}$ using (\ref{eq:hll_fluxes}) with left and 
right states given, respectively, by $\A{U}_i$ and $\A{U}_{i+1}$.
 
For the second order scheme, the input to the Riemann 
problem are the states  
\begin{equation}\label{eq:LRstates}
 \begin{array}{c}
 \DS  \A{V}^{n+\HALF}_{i+\HALF,L} = \A{V}^{n+\HALF}_i + 
                           \frac{\delta\A{V}^n_i}{2}
  \, , \\ \noalign{\bigskip}
 \DS  \A{V}^{n+\HALF}_{i+\HALF,R} = \A{V}^{n+\HALF}_{i+1} - 
                           \frac{\delta\A{V}^n_{i+1}}{2} \,,
 \end{array}
\end{equation}
where $\A{V}^{n+\HALF}_i$ follows from 
a simple Hancock predictor step, 
\begin{equation}\label{eq:hancock_pred}
 \A{U}^{n+\HALF}_i = \A{U}^n_i - \frac{\Delta t^n}{2\Delta x_i}\left[
        \A{F}\left(\A{V}^n_{i+\HALF,L}\right) -
        \A{F}\left(\A{V}^n_{i-\HALF,R}\right)\right]\,,
\end{equation}
with $\A{V}^n_{i+\HALF,L}$ and $\A{V}^n_{i-\HALF,R}$ computed 
from (\ref{eq:LRstates}) by replacing $\A{V}^{n+\HALF}$ with $\A{V}^{n}$.

The $\delta \A{V}$'s appearing in equation 
(\ref{eq:LRstates}) are computed at the beginning of the time step 
using the fourth-order limited slopes \citep{Colella85, Saltzman94}:
\begin{equation}
  \delta\A{V}_i = s_i \min\left(\left|\frac{4}{3} \Delta_0\A{V}_i - 
               \frac{\bar{\delta}\A{V}_{i+1} + \bar{\delta}\A{V}_{i-1}}{6}\right|,
               \Delta_l \A{V}_i\right)\;,
\end{equation}  
where 
\begin{equation}
 \Delta_l\A{V}_i = \alpha \min\left(|\Delta\A{V}_i|, |\Delta\A{V}_{i-1}|\right)\;, 
\end{equation}
and $\bar{\delta}\A{V}_i$ are the second-order slopes
\begin{equation}
 \bar{\delta}\A{V}_i = s_i \min\left(\Delta_l\A{V}_i, |\Delta_0\A{V}_i|\right)    \;,
\end{equation}
\begin{equation}
  \DS  \Delta\A{V}_i = \A{V}_{i+1} - \A{V}_i \;,\quad 
  \Delta_0\A{V}_i = \frac{\A{V}_{i+1} - \A{V}_{i-1}}{2}   \;, 
\end{equation}
\begin{equation}
  \DS s_i = \frac{\sign(\Delta\A{V}_i) + \sign(\Delta\A{V}_{i-1})}{2} \;.
\end{equation}
The parameter $\alpha\in[1,2]$ adjusts the limiter compression,
with $\alpha = 2$ ($\alpha = 1$) yielding a more (less) compressive
limiter.
Notice that, although the use of fourth-order slopes attains sharper representations of 
discontinuities, the scheme retains global second-order spatial accuracy.

We do not make use of any artificial steepening 
algorithm to enhance resolution across a contact wave 
\citep{MM96, SFIM04, MPB05} in order
to highlight the intrinsic capabilities of our new HLLC solver.
In the one-dimensional tests, the computational domain is the 
interval $[0,1]$ and the compression parameter is $\alpha = 2$.
In two dimensions we set $\alpha = 2$, $\alpha = 1.25$ and 
$\alpha = 1$ for density, velocities and pressure, respectively.
Additional shock flattening, computed as in \citep{MM96}, is
used in \S\ref{sec:p1} and \S\ref{sec:p5} to 
prevent spurious numerical oscillations. 
Outflow boundary conditions are set in problem
$1$--$4$.

Multidimensional integration is achieved via Strang directional splitting  
\citep{Strang68}, that is, by successively applying one-dimensional operators
in reverse order from one time step to the next one,
i.e. $\A{U}^{n+2} = {\cal L}_x{\cal L}_y \A{U}^{n+1}$ and
$\A{U}^{n+1} = {\cal L}_y{\cal L}_x \A{U}^{n}$.
Here ${\cal L}_x$ is the operator corresponding to the conservative 
update (\ref{eq:update}) (and similarly for ${\cal L}_y$).
The same time increment $\Delta t$ should be used for two consecutive
time steps.

%%%%%%%%%%%%%%%%%%%%%%%%%%%%%%%%%%%%%%%%%%%%%%%%%%%%%%%%%%%%%%%%%%%%%
\subsection{Problem 1}\label{sec:p1}
%
%
%%%%%%%%%%%%%%%%%%%%%%%%%%%%%%%%%%%%%%%%%%%%%%%%%%%%%%%%%%%%%%%%%%%%%

The first test consists in a Riemann problem with initial data
\begin{equation}
 \big(\rho, v_x, p\big)  = \left\{\begin{array}{cc}
   \big(1, 0.9, 1\big)  & \; \textrm{for} \quad x < 0.5  \,, \\ \noalign{\medskip} 
   \big(1, 0, 10 \big)  & \; \textrm{for} \quad x > 0.5  \,.
  \end{array}\right.
\end{equation}
Integration is carried with $CFL = 0.8$ until $t = 0.4$ and
an ideal equation of state with $\Gamma = 4/3$ is used.
The breakup of the discontinuity results in the formation
of two shock waves separated by a contact discontinuity.

\begin{figure}
 \includegraphics[width=80mm]{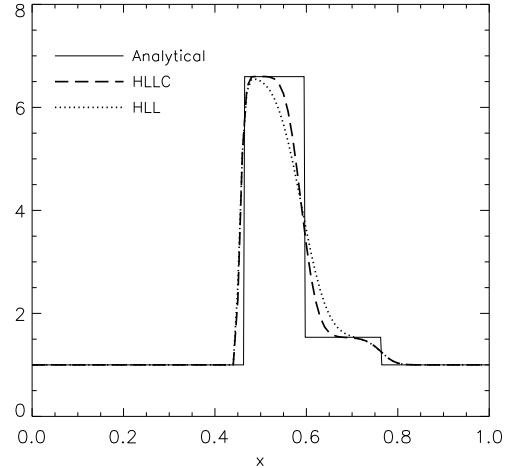}
 \caption{Comparison between the HLL (dotted line) and HLLC 
         (dashed line) Riemann solvers for problem $1$ at $t=0.4$. Only the 
          density profiles are shown. Computations were 
          performed with the first-order scheme on $100$ computational zones 
          with CFL = 0.8. 
          The solid line gives the analytical solution. 
          The major difference between the two approaches is the 
          resolution of the contact wave.}
 \label{fig:flat_1}
\end{figure}

In Fig. \ref{fig:flat_1} we plot the analytical solution
for the rest mass density together with the profiles obtained  
with the first-order HLLC and HLL schemes on $100$ uniform 
computational zones.
The two integrations behave similarly near the shock waves, 
but differ in the ability to resolve the contact discontinuity. 
As expected, the HLLC scheme yields a sharper representation
of the latter, whereas the HLL solver retains a 
more diffusive character.

\begin{figure}
 \includegraphics[width=80mm]{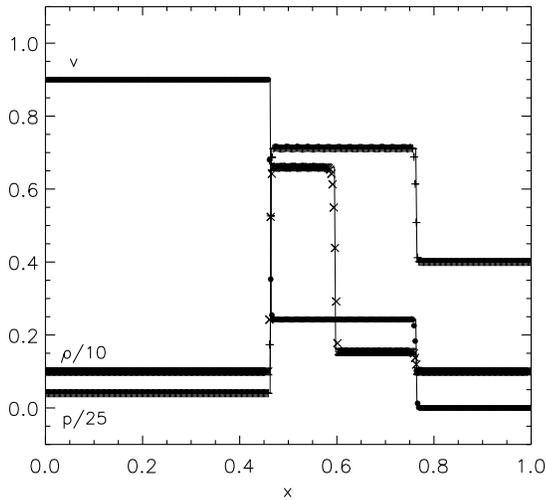}
 \caption{Numerical solutions obtained with the second-order HLLC 
          scheme applied to problem $1$. 
          The solid line represents the analytical solution, while 
          computed profiles of density (crosses), pressure (plus signs) and 
          velocity (filled circles) are shown on $400$ computational 
          zones at $t=0.4$. A CFL number of $0.8$ was used.}
 \label{fig:sod_1}
\end{figure}

The $L_1$ norm errors of density are shown in the top-left
panel of Fig. \ref{fig:res_flat} for different resolutions.
For the sake of comparison, computations have also been performed 
using the more sophisticated exact Riemann solver 
described in the one-dimensional code by \citet{MM96}.
The errors obtained with the present HLLC scheme and the exact Riemann 
solver are comparable at low resolution ($\sim 15.3\%$ and 
$\sim 13.6 \%$ respectively on $100$ points) and become nearly 
identical as the number of points increases.
Conversely, the errors computed with the relativistic 
HLL scheme are bigger ($\sim 22.2\%$ on $100$ points) and 
show that almost twice the resolution is needed
to achieve the same accuracy obtained with the HLLC or the 
exact solver.
    
Fig. \ref{fig:sod_1} shows the results computed  
with the second-order HLLC scheme on $400$ zones, at the same time.
The exact profiles for density, velocity and pressure are 
plotted as solid lines.
Additional slope flattening \citep{MM96} has been used to 
reduce the spurious numerical oscillations observed behind the shock
front.
All discontinuities are adequately captured and resolved on 
few computational cells, $\sim 3$ for the shocks 
and $\sim 4-5$ for the contact discontinuity 
(contrary to $\sim 7$ when the HLL solver is employed). 

The error in $L_1$ norm is $\sim 2.3\%$ for $400$ grid zones 
and it has been computed at different resolutions 
using the HLL, HLLC and exact Riemann solver, see Fig. 
\ref{fig:res_2nd}.
Not surprisingly, the second-order interpolation considerably reduces 
the errors and higher convergence rates are expected for all schemes.
Nevertheless, the three solvers mostly differ in the resolution at the 
contact discontinuity and, for $n\ge800$ grid 
points, the HLLC and exact Riemann are practically indistinguishable,   
while at the maximum resolution employed ($3200$ zones), the error
computed with the HLL scheme is still $\sim 20\%$ bigger.

%%%%%%%%%%%%%%%%%%%%%%%%%%%%%%%%%%%%%%%%%%%%%%%%%%%%%%%%%%%%%%%%%%%%%
\subsection{Problem 2}\label{sec:p2}
%
%
%%%%%%%%%%%%%%%%%%%%%%%%%%%%%%%%%%%%%%%%%%%%%%%%%%%%%%%%%%%%%%%%%%%%%

In the second test, we prescribe the initial condition
\begin{equation}
 \big(\rho, v_x, p\big)  = \left\{\begin{array}{cc}
   \big(1, -0.6, 10\big)  & \; \textrm{for} \quad x < 0.5  \,, \\ \noalign{\medskip} 
   \big(10, 0.5, 20\big)  & \; \textrm{for} \quad x > 0.5  \,,
  \end{array}\right.
\end{equation}
with an ideal equation of state with $\Gamma = 5/3$. 
Integration stops at $t = 0.4$ and CFL = 0.8 has been used
in the integration.
The initial discontinuity evolves into left-going and 
right-going rarefaction waves
with a contact discontinuity in the middle.
 
\begin{figure}
 \includegraphics[width=80mm]{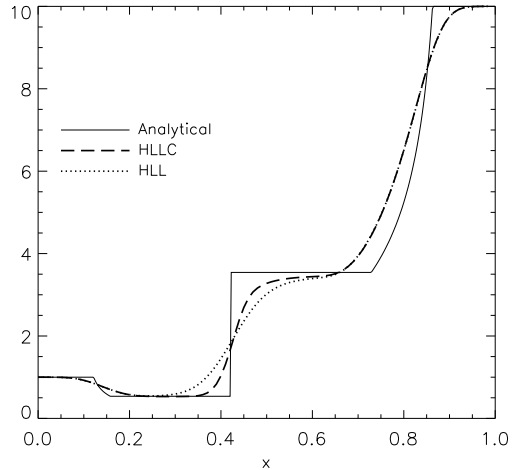}
 \caption{Computed density profiles for the first-order HLL 
          (dotted line) and HLLC (dashed line) schemes for problem $2$
          at $t=0.4$.
          We used $100$ grid points and CFL number of 0.8. 
          The analytical solution is plotted as a solid line. 
          As expected, the HLLC solver behaves quantitatively 
          better than the HLL scheme across the contact wave.}
 \label{fig:flat_2}
\end{figure}
\begin{figure}
 \includegraphics[width=80mm]{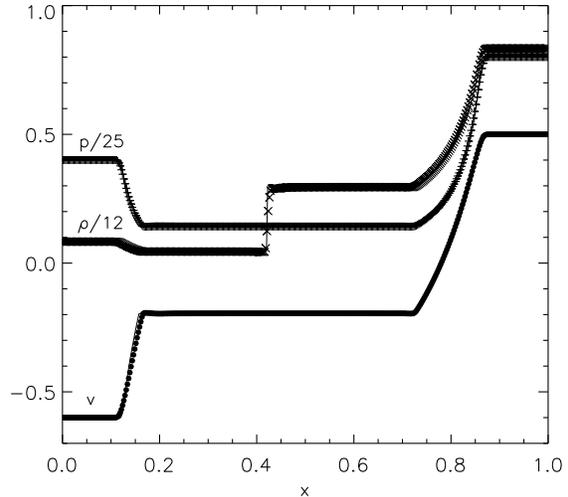}
 \caption{Second-order HLLC scheme applied to problem 2 at 
          $t = 0.4$ on $400$ computational zones with CFL = 0.8.
          As in Fig. \ref{fig:sod_1}, profiles for density, pressure
          and velocity are plotted using crosses, plus signs and filled 
          circles. The contact wave is the only discontinuity in the solution
          and is clearly visible at $x\approx 0.4$.}
 \label{fig:sod_2}
\end{figure}

Results for the first-order HLL and HLLC schemes 
on a $100$-point uniform grid are shown in Fig. 
\ref{fig:flat_2}.
Again, notice the sharper resolution of the HLLC scheme
in proximity of the contact wave.
The smooth rarefaction waves are equally resolved by both schemes.
 
The behavior of the solution under grid resolution effects is 
described in the top-right panel of Fig. \ref{fig:res_flat}.
Since the only discontinuity in the problem is the contact wave, 
the $L_1$ norm reflects mostly the different resolution 
across the discontinuity.
The HLLC and the exact solver perform nearly 
identically, while the HLL exhibits a slightly slower convergence rate. 
At the maximum resolution, the error in the HLL scheme is 
$\sim 4.3\%$ to be compared to the $\sim 3.0\%$ and $3.1\%$ errors 
obtained from the other two Riemann solvers. 

These differences are again reduced in the second-order 
HLLC scheme, Fig. \ref{fig:sod_2}, for which the convergence 
rates are similar, as shown in the top-right panel of Fig. 
\ref{fig:res_2nd}.

%%%%%%%%%%%%%%%%%%%%%%%%%%%%%%%%%%%%%%%%%%%%%%%%%%%%%%%%%%%%%%%%%%%%%
\subsection{Problem 3}\label{sec:p3}
%
%
%%%%%%%%%%%%%%%%%%%%%%%%%%%%%%%%%%%%%%%%%%%%%%%%%%%%%%%%%%%%%%%%%%%%%

\begin{figure}
\includegraphics[width=80mm]{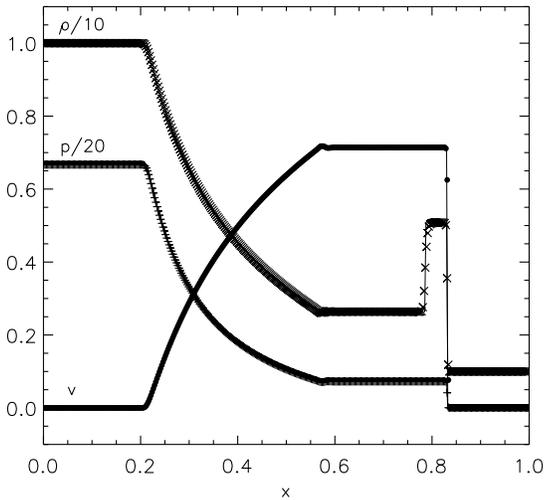}
 \caption{Results of the second-order HLLC scheme applied to shock tube 
          problem 3 at $t = 0.4$ on $400$ computational zones. 
          Integration has been carried with CFL = 0.8.
          The solution is comprised of a left-going rarefaction wave, 
          a right-going contact and shock wave moving close to 
          each other.}
 \label{fig:sod_3}
\end{figure}

The initial condition for this test is
\begin{equation}
 \big(\rho, v_x, p\big)  = \left\{\begin{array}{cc}
   \big(10, 0, 40/3\big)  & \; \textrm{for} \quad x < 0.5  \,, \\  \noalign{\medskip} 
   \big(1,  0, 0\big)  & \; \textrm{for} \quad x > 0.5   \,,
  \end{array}\right.
\end{equation}
with $\Gamma = 5/3$.
For numerical reasons, the pressure in the left state has been 
set equal to a small value, $p=2/3\cdot 10^{-6}$.
Integration is carried with CFL = 0.8 on $400$ grid points;
the final integration time is $t = 0.4$.
The initial configuration results in a mildly relativistic
blast wave, with a maximum Lorentz factor $\gamma_{\max}\sim 1.4$.
The Riemann fan consists of a rarefaction wave moving
to the left, a shock wave adjacent to a contact discontinuity, 
both moving to the right, see Fig. \ref{fig:sod_3}.  
 
Our HLLC scheme is able to capture discontinuities properly;
in particular, the shock is resolved within 2-3 zones and 
the contact discontinuity smears out over 4-5 
zones.  We remind again that the interpolation algorithm does not 
make use of additional artificial compression to enhance 
resolution across the contact wave, as in \cite{MM96}. 
Moreover, we repeated the test also with the exact Riemann solver 
and did not find any noticeable difference.
In addition and contrary to the previous two test problems, we did not 
find strong differences between our HLLC method and the HLL scheme. 
Resolution effects are given in the bottom left panels of Fig.
\ref{fig:res_flat} and \ref{fig:res_2nd} for the first-order and second
order schemes, respectively. As one can see, the solutions 
computed with the HLL, HLLC and exact solvers behave nearly in the same
way, with the $L_1$ norm errors being different by less than $1\%$ at
low resolution and becoming identical for $n \ge 800$ grid points. 
We believe that this might be due to the proximity of the contact 
and shock waves.
The quality of our results is, however, similar and comparable to
those obtained in previous studies. 

%%%%%%%%%%%%%%%%%%%%%%%%%%%%%%%%%%%%%%%%%%%%%%%%%%%%%%%%%%%%%%%%%%%%%
\subsection{Problem 4}\label{sec:p4}
%
%
%%%%%%%%%%%%%%%%%%%%%%%%%%%%%%%%%%%%%%%%%%%%%%%%%%%%%%%%%%%%%%%%%%%%%

\begin{figure}
\includegraphics[width=80mm]{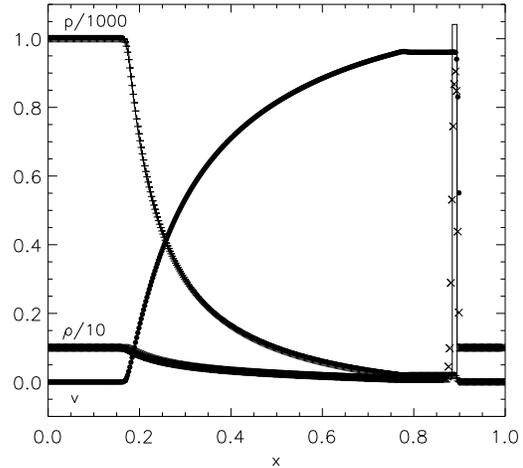}
 \caption{Computed profiles of density, pressure 
          and velocity for the second-order HLLC 
          scheme applied to blast wave problem (problem $4$). 
          Integration has been carried with $CFL = 0.8$ on $400$ uniform 
          zones until $t=0.4$.
          The configuration is similar to that of problem $3$, but
          the shock and the contact waves are now much closer
          to each other.}
 \label{fig:sod_4}
\end{figure}

In the fourth shock-tube, we prescribe the following initial discontinuity 
\begin{equation}
 \big(\rho, v_x, p\big)  = \left\{\begin{array}{cc}
   \big(1, 0, 10^3\big)  & \; \textrm{for} \quad x < 0.5  \,,  \\ \noalign{\medskip} 
   \big(1, 0, 10^{-2}\big)  & \; \textrm{for} \quad x > 0.5  \,.
  \end{array}\right.
\end{equation}
Again we adopt an ideal equation of state with $\Gamma = 5/3$.
The resulting pattern is similar to that of problem $3$, but the 
specific enthalpy in the left state is much greater than unity, 
thus resulting in a more thermodynamically relativistic configuration.
The solution computed with the second-order scheme at $t = 0.4$ is 
shown in Fig. \ref{fig:sod_4} on $400$ computational zones and 
CFL=0.8.

The high pressure jump produces a strong shock wave and a contact 
discontinuity very close to each other, moving to the right at 
almost the same speeds. 
The higher compression in the shell is due to a more 
pronounced relativistic length-contraction effect caused by a 
higher Lorentz factor, $\gamma_{\max}\sim 3.7$.
The smaller thickness of the shell between the shock and the contact wave
makes this test numerically challenging and particularly demanding in terms
of resolution.

Our relativistic HLLC scheme is able to reproduce the solution
within a satisfactory agreement, even without using a contact 
steepening algorithm. 
The absolute global error in density is $6.5\%$ and the density peak 
in the thin shell achieves $\sim 81.6\%$ of the exact value.
Our results are therefore similar to previous ones proposed in 
literature.

\begin{figure}
\includegraphics[width=80mm]{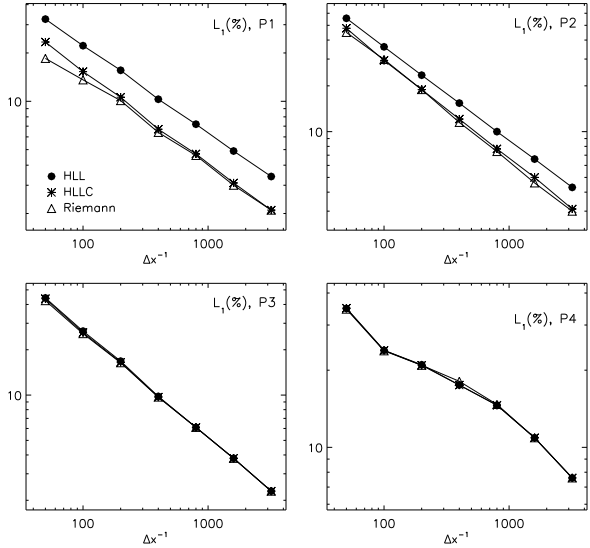}
 \caption{Grid resolution studies for the first-order schemes using 
          the HLL (filled circles), HLLC (crosses) and exact Riemann
          solvers (triangles).
          Results are shown for problem $1$ (P1,
          top left), problem $2$ (P2 top right), problem $3$ (P3,
          bottom left) and problem $4$ (P4, bottom right).
          Computations have been performed on $50$, $100$,
          $200$, $400$, $800$, $1600$ and $3200$ grid zones 
          with the same CFL number (0.8) for all runs. 
          Notice that both the HLLC and exact solvers perform 
          better than the HLL scheme in problem 
          $1$ and $2$, while all schemes yield nearly identical
          results in problem $3$ and $4$.}
\label{fig:res_flat}
\end{figure}

\begin{figure}
\includegraphics[width=80mm]{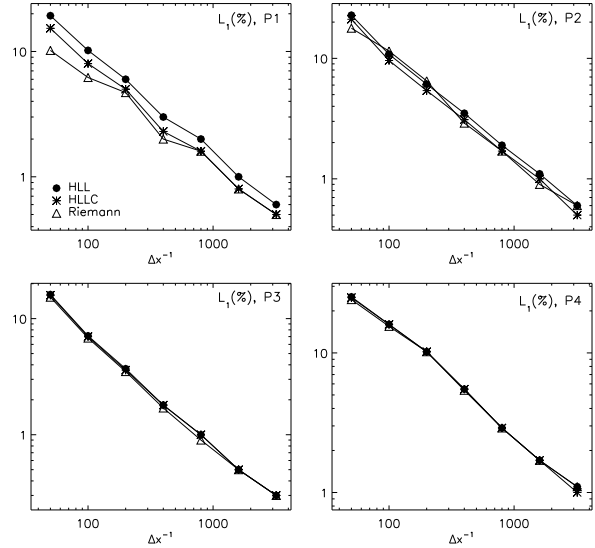}
 \caption{Grid resolution studies for the second-order  
          schemes. The notations are the same ones used 
          in Fig. \ref{fig:res_flat}. 
          Computations were obtained using the same CFL number 
          (0.8) for all cases.}
\label{fig:res_2nd}
\end{figure}

It should also be pointed out that, similarly to problem 3, we did
not find any improvement in the solution by switching to the 
exact Riemann solver or using the relativistic HLL scheme.
This is confirmed by the resolution study carried for the first and
second-order schemes (bottom right panels in Fig. \ref{fig:res_flat}
and \ref{fig:res_2nd}).
Again, we suggest that the ability to capture the discontinuities 
relies on the interpolation properties of the algorithm
and has a weaker dependence on the Riemann solver for this
particular class of problems.

%%%%%%%%%%%%%%%%%%%%%%%%%%%%%%%%%%%%%%%%%%%%%%%%%%%%%%%%%%%%%%%%%%%%%
\subsection{Relativistic Planar Shock Reflection}\label{sec:p5}
%
%
%%%%%%%%%%%%%%%%%%%%%%%%%%%%%%%%%%%%%%%%%%%%%%%%%%%%%%%%%%%%%%%%%%%%%

The initial configuration for this test problem consists in a 
cold ($p = 0$), uniform ($\rho = 1$) flow impinging on a wall 
located at $x=0$. The flow has constant inflow velocity $v_{in}$ 
and the reflection results in the formation of a strong 
shock wave. For $t>0$ the shock propagates upstream and
the solution has an analytic form given by \citep{MM96}:
\begin{equation}
  \rho(r,t) = \left\{ \begin{array}{cc}
\DS   1       &  \; \textrm{for}\quad r > v_st \, , \\  \noalign{\medskip}
\DS   \sigma  &  \; \textrm{for}\quad r < v_st \, ,
  \end{array}\right.
\end{equation}
where 
\begin{equation}
  \sigma = \frac{\Gamma + 1 + \Gamma\left(\gamma_{in} - 1\right)}{\Gamma - 1}\,,
\quad
  v_s = (\Gamma - 1)\frac{\gamma_{in} |v_{in}|}{\gamma_{in} + 1} \,,
\end{equation}
are the compression ratio and the shock velocity, respectively.
Behind the shock wave ($r < v_st$), the gas is at rest (i.e. $v=0$) and the 
pressure has the constant value $\rho(r,t)(\gamma_{in} - 1)(\Gamma - 1)$.
Conversely, in front of the shock all of the energy is kinetic and thus $p=0$,  
$v = v_{in}$.

For numerical reasons, pressure has been initialized to a small finite 
value, $p = \epsilon(\Gamma - 1)$, with $\epsilon = 10^{-10}$ and 
$\Gamma = 4/3$.
The computational domain is covered with 
$100$ computational zones and the initial inflow velocity 
is $v_{in} = -0.99999$ corresponding to a Lorentz factor 
$\gamma_{in}\sim 224$. Integration is carried with CFL =0.4. 

\begin{figure}
\includegraphics[width=80mm]{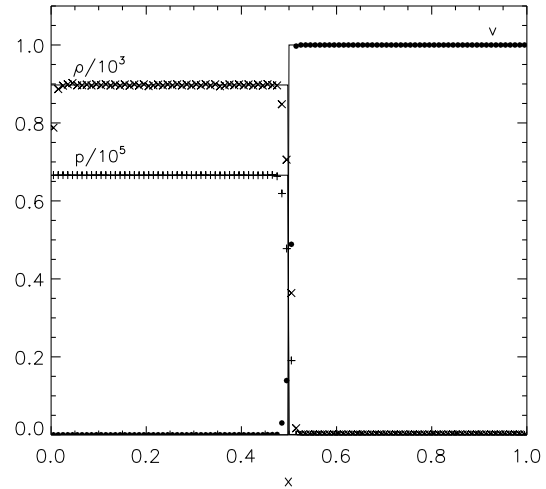}
 \caption{Relativistic planar shock reflection test with CFL = 0.4 and 
          $100$ grid points. Results are shown at $t=1.5$ for the
          second-order HLLC scheme. The cold supersonic flow 
          enters at $x=1$ and a reflective boundary condition
          is imposed on the left, at $x=0$.
          The reflected shock is located at $x=0.5$.}
\label{fig:rpsr}
\end{figure}

Fig. \ref{fig:rpsr} shows the solution at $t= 1.5$, after
the shock has propagated $\Delta x_s \approx 0.5$ from the 
wall. 
The relative global errors (defined as $\epsilon(L_1)/\sum_i \rho_{ex}(x_i)\Delta x_i$) 
for density, velocity and pressure are, respectively, 
$1.8\%$, $1.4\%$ and $1.4\%$.
This result is in excellent quantitative agreement with
the numerical solutions obtained by other authors 
\citep{MMIMD92, MMFIM97, Aloy99, dZB02}.
In this test, similarly to problem $1$, shock flattening 
was employed to prevent numerical oscillations.

Also, from the same figure, we notice that our solver 
suffers from the wall-heating phenomenon,  a common pathology 
in modern Godunov-type schemes.
The degree of this pathology is higher than the HLL scheme 
but less than the exact Riemann solver.
We also point out that the problem may be partially mitigated 
by a proper fine-tunings of the parameters involved 
in the reconstruction and steepening algorithms. 
However, we did not follow that approach in the present work.

%%%%%%%%%%%%%%%%%%%%%%%%%%%%%%%%%%%%%%%%%%%%%%%%%%%%%%%%%%%%%%%%%%%%%
\subsection{Two-Dimensional Riemann Problem}\label{sec:p6}
%
%
%%%%%%%%%%%%%%%%%%%%%%%%%%%%%%%%%%%%%%%%%%%%%%%%%%%%%%%%%%%%%%%%%%%%%

Two dimensional Riemann problems involve the interactions  
of four elementary waves (either shocks, rarefactions,
and contact discontinuities) initially separating four constant states.
They have been formulated by \citet{SCG93}, \citet{LL98} in the 
context of classical hydrodynamics.
Here we consider a relativistic extension, initially proposed 
by \citet{dZB02}, where the initial configuration involves two 
shocks and two tangential discontinuities.

The domain is the square $[-1,1]\times[-1,1]$ covered with $400^2$
computational zones. The four quadrants $NE$ ($x,y>0$), 
$NW$ ($x < 0 < y$), $SW$ ($x,y < 0$), $SE$ ($y < 0 < x$) divide 
the square into four constant-state regions:
\begin{equation}\label{eq:riemann2d}
 \big(\rho, v_x, v_y, p\big) = \left\{\begin{array}{ll}
 \big(0.1,   0, 0   , 0.01\big) & \;\textrm{for}\quad x,y >0    \,, \\ \noalign{\medskip}
 \big(0.1,0.99, 0   ,    1\big) & \;\textrm{for}\quad x < 0 < y \,, \\ \noalign{\medskip}
 \big(0.5,   0, 0   ,    1\big) & \;\textrm{for}\quad x,y < 0   \,, \\ \noalign{\medskip}
 \big(0.1,   0, 0.99,    1\big) & \;\textrm{for}\quad y < 0 < x \,.\\ \noalign{\medskip}
\end{array}\right.
\end{equation}
We use an ideal equation of state with $\Gamma = 5/3$. The integration 
is carried out with CFL=0.4 till $t=0.8$.

Notice that the initial condition (\ref{eq:riemann2d}) does not 
exactly prescribe two simple shock waves
at the $NW-NE$ and $SE-NE$ interface. The correct version of this
problem has been considered by \citet{MPB05}. For the sake of comparison,
however, we chose to adopt the same initial condition as in \cite{dZB02}.

\begin{figure}
\includegraphics[width=85mm]{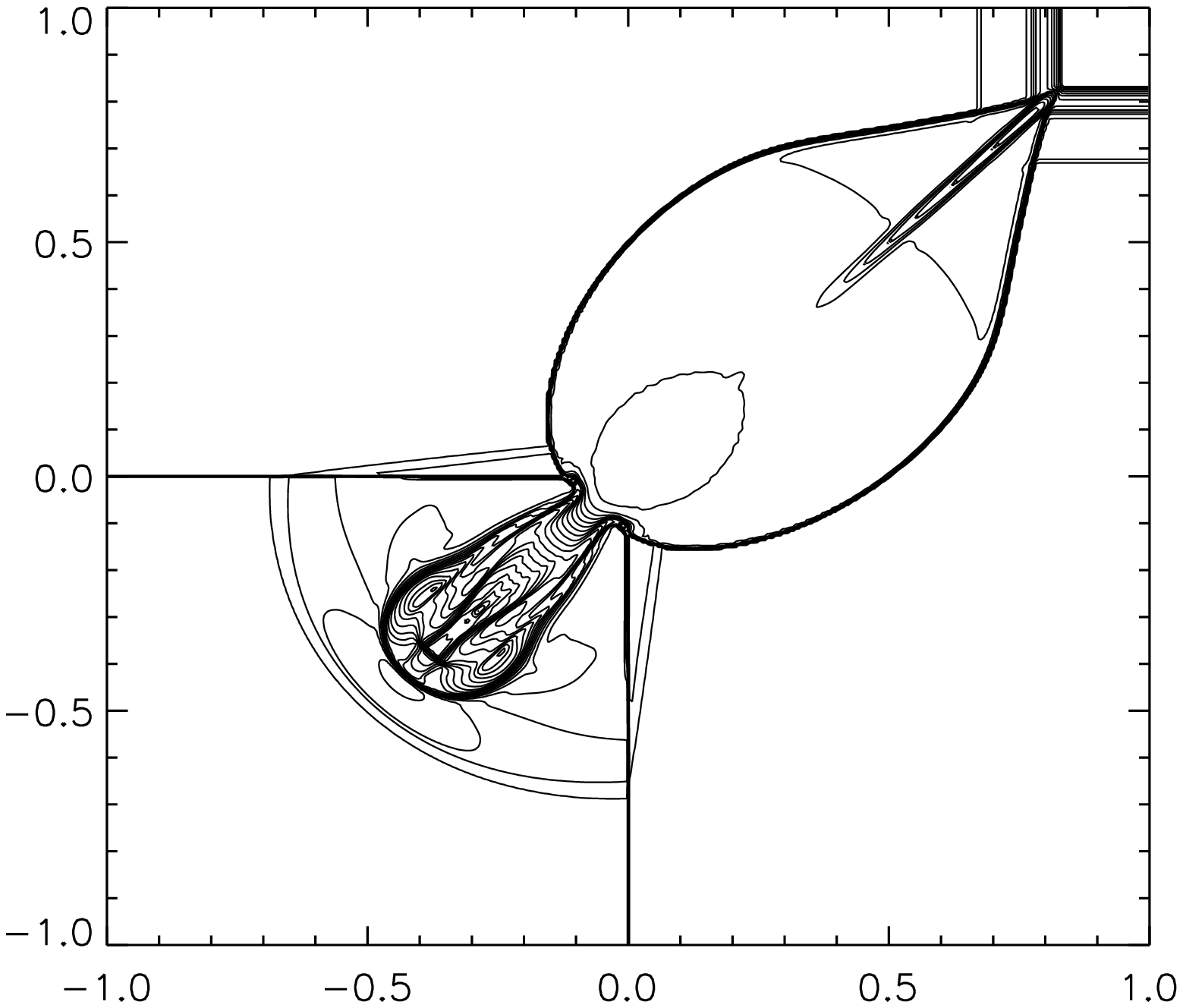}
\includegraphics[width=85mm]{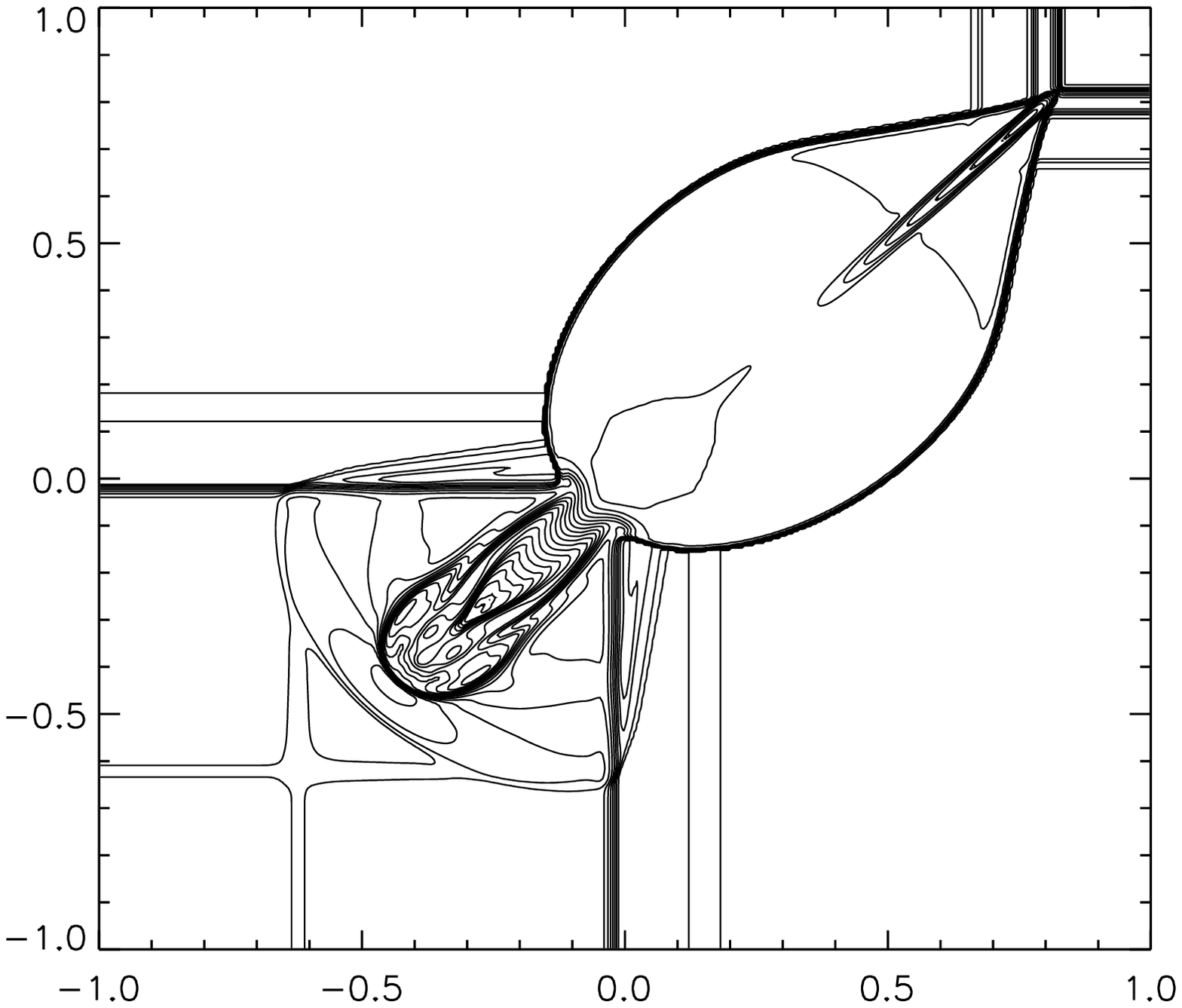}
 \caption{Density logarithms for the two-dimensional Riemann problem 
          on $400^2$ zones at $t = 0.8$; the top (bottom) panel shows
          the results obtained with the second-order HLLC (HLL) scheme. 
          Thirty equally spaced contours are shown.
          Integration has been performed with CFL = 0.4.
          Curved transmitted shocks are visible in the upper right
          portion of the domain. The drop-shaped region in the
          lower left portion is bounded by two tangential 
          discontinuities. The HLL (bottom panel) shows additional 
          numerical diffusion which degenerate into spurious 
          waves propagating along the main axis.}
\label{fig:riemann2d}
\end{figure}

The top and bottom panels in Fig. \ref{fig:riemann2d} show, 
respectively, the solutions computed with the HLLC and HLL solvers.
The breakup of the initial discontinuity results in two equal-strength 
curved shock fronts propagating from regions $NW$ and $SE$ into the 
upper right portion of the domain ($NE$), top panel of Fig. \ref{fig:riemann2d}.
Region $SW$ is bounded by two tangential discontinuities and a jet-like
structure emerges along the main diagonal.

The initial steady tangential discontinuities, located at the 
$W$ and $S$ interfaces, remain automatically
sharp in the HLLC formulation, since they are exactly captured by the 
approximate Riemann solver. The same results has also been shown by 
\citet{MPB05} who used a two-shock iterative nonlinear solver.
We emphasize that this is property pertains to the Riemann solver 
itself and does not depend on the interpolation algorithm.
Indeed the same result holds when the first-order scheme 
is employed.
This feature is absent from the HLL formulation, where tangential 
discontinuities spread along the cartesian axis due to extra 
numerical diffusion.
This behavior is manifestly evident in the grid-aligned spurious waves 
visible in the bottom panel of Fig. \ref{fig:riemann2d},
see also \cite{dZB02, SFIM04}.

%%%%%%%%%%%%%%%%%%%%%%%%%%%%%%%%%%%%%%%%%%%%%%%%%%%%%%%%%%%%%%%%%%%%%
\subsection{Axisymmetric Relativistic Jet}\label{sec:p7}
%
%
%%%%%%%%%%%%%%%%%%%%%%%%%%%%%%%%%%%%%%%%%%%%%%%%%%%%%%%%%%%%%%%%%%%%%

Finally, as an example of an astrophysical application, we consider
the propagation of a light, axisymmetric relativistic jet in 2-D 
cylindrical geometry. For the sake of
comparison, the  parameters of the simulation are the same used by
\citet{dZB02} and by \citet{SFIM04}. 
The initial condition is prescribed as
\begin{equation}
 \big(\rho, v_r, v_z, p\big) = \left\{\begin{array}{ll}
 \big(0.1,\,  0,\, 0.99,\, 10^{-2}\big) & \;\textrm{for}\quad r,z < 1    \,, \\ \noalign{\medskip}
 \big(10, \,  0,\, 0,   \, 10^{-2}\big) & \;\textrm{otherwise}           \,. \\ \noalign{\medskip}
\end{array}\right.
\end{equation}
The jet is pressure matched and its internal 
relativistic Mach number is 17.1. We use an ideal equation of state 
with $\Gamma = 5/3$ both for the jet and the ambient medium. 
The computational domain covers the
region $0 \leq r \leq 12$, $0 \leq z \leq 35$, with  $240
\times 700$ grid points, so that we have 20 cells per jet radius. 
At the symmetry axis, $r=0$, we impose reflecting boundary conditions;
outflow boundary conditions are set everywhere else, except at the inlet
region where we keep the beam constant. 
The CFL number is $0.5$ and
the jet evolution is followed until $t = 80$. 

\begin{figure}
\includegraphics[width=85mm]{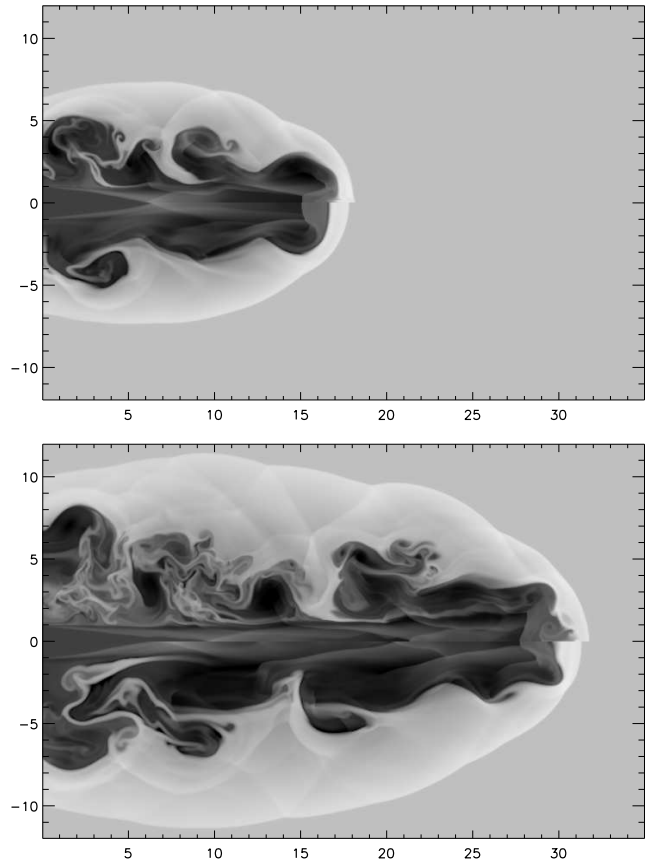}
 \caption{Gray scale images of the rest-mass density logarithm for the
          relativistic jet simulation at $t=40$ 
          (top panel) and $t = 80$ (bottom panel). 
          In each panel, the HLLC (HLL) solver has been used 
          for the upper (lower) portion of the image. 
          The resolution is $20$ points per jet radius, 
          corresponding to grid size of $240\times 700$ 
          computational zones. 
          Integration has been carried with CFL = 0.5}
\label{fig:jet}
\end{figure}

For the sake of comparison, we have also carried the simulation 
using the relativistic HLL solver. 
Figure \ref{fig:jet} shows two snapshots of the rest mass 
density at times $t = 40$ and $t = 80$. 
The upper-half of each panel refers to the HLLC integration, whereas
the lower portion displays the result obtained with the HLL
scheme.
We see that all the structural features characteristic of jet
propagation can be clearly identified, with good resolution of shock
waves and contact discontinuities. 
It is clear from the figures that the HLLC integration shows 
a significantly greater amount of small scale structure, that is
not visible in the HLL results. This is due to the larger 
numerical diffusion introduced by the latter in subsonic regions 
which prevent sharp resolution of shear and tangential waves.

The average advance speed of the jet head is $\sim 0.39$
(to be compared with a onedimensional theoretical
estimate of $0.44$; \citet{MMFIM97}). Moreover, we can observe the absence
of the carbuncle problem, that usually appears as an extended nose in
front of the jet, on the axis \citep{Quirk94}.

%%%%%%%%%%%%%%%%%%%%%%%%%%%%%%%%%%%%%%%%%%%%%%%%%%%%%%%%%%%%%%%
\section{Efficiency Comparison}
%
%
%
%
%%%%%%%%%%%%%%%%%%%%%%%%%%%%%%%%%%%%%%%%%%%%%%%%%%%%%%%%%%%%%%%

Previous numerical tests have shown that the
quality of solution achieved with the HLLC scheme can be 
competitive with more complex exact or iterative non-linear 
Riemann solvers, see for example \cite{MM96}, \cite{MPB05}.
Another aspect which plays in favour of the HLLC formalism is
the computational efficiency, particularly crucial in long-term 
simulations in two and three dimensions.

Table \ref{tab:tci} gives the normalized CPU time required 
by the HLL, HLLC and approximate two-shock nonlinear Riemann solvers 
\cite[for the latter see][]{MPB05}. 
All three solvers are available in the C author's code and
have been written performing similar degree of optimizations.
On the contrary, the FORTRAN code for the exact solution 
to the Riemann problem, available from \cite{MM03}, was 
found to be more than a factor of $7$ slower than the HLL solver. 
We believe that this might be due to a lower degree
of optimization and by the time consuming numerical integration
across the rarefaction fan \citep{PMM00}.
\begin{table}\begin{center}
\begin{tabular}{c|c|c|c|c}\hline
  Test   & \#\,zones  & HLL   & HLLC   & Riemann  \\ \hline\hline
\# $1$   & $4000$  & $1$   & $1.05$ & $1.37$     \\  \hline
\# $2$   & $4000$  & $1$   & $1.07$ & $1.34$     \\ \hline
\# $3$   & $4000$  & $1$   & $1.06$ & $1.30$     \\ \hline
\# $4$   & $4000$  & $1$   & $1.04$ & $1.44$     \\ \hline
   R2-D  & $400^2$ & $1$   & $1.03$ & $1.30$    \\  \hline\hline
\end{tabular}
\caption{Normalized CPU time per integration step for the first four 
         one-dimensional shock tubes and the 2-D Riemann
         problem (R2-D) considered in \S\ref{sec:p6}. 
         The two rightmost columns 
         give the average computing time for the HLLC and two-shock 
         nonlinear Riemann solvers normalized to the HLL time
         (third column).
         All runs were produced using the first-order scheme 
         with $CFL = 0.8$.}
\label{tab:tci}
\end{center}\end{table}
For illustrative purposes, we consider the first four one-dimensional 
tests and the two-dimensional Riemann problem.
Integrations have been carried using the first-order scheme 
on $4000$ and $400^2$ zones, respectively. 
No optimization flags were used during the compilation.

From the table, one can easily conclude that the HLLC scheme 
requires little additional costs with respect to the HLL approach
(between $4\%$ and $7\%$ in one dimension),
while the iterative nonlinear solver is certainly more
expensive, being by a factor of more than $30\%$ slower.

In making the comparison, however, one should keep in mind that 
HLL-type solvers are iteration-free since  
the underlying algorithms always require a fixed number of 
operations, regardless of the initial condition.
On the contrary, iterative nonlinear Riemann solvers 
have a certain degree of adaptability since the number
of iterations to achieve convergence depends on the 
strength of the discontinuity at the zone interface. 
In smooth regions of the flow, for example, fewer iterations
are usually needed.
This explains why the fourth test problems is particularly
time consuming, since a stronger discontinuity is 
involved.   

In this respect, a direct comparison between different
Riemann solvers becomes problem-dependent and can be used only
as an order-of-magnitude estimate. 
Conversely, we do not expect the HLLC/HLL efficiency 
ratio to change with increasing complexity of the flow patterns. 
For this reason, we believe that for problems involving rich and
complex structures the trade-off between computational 
efficiency and quality of results is certainly worth the effort.

%%%%%%%%%%%%%%%%%%%%%%%%%%%%%%%%%%%%%%%%%%%%%%%%%%%%%%%%%%%%%%%
\section{Conclusions}
%
%
%
%
%%%%%%%%%%%%%%%%%%%%%%%%%%%%%%%%%%%%%%%%%%%%%%%%%%%%%%%%%%%%%%%

We have presented, for the first time, an extension 
of the HLLC scheme by \citet{TSS94} to relativistic gas dynamics. 
The solver is robust, computationally efficient and 
complete, in that it considers the full wave structure 
in the solution to the Riemann problem.
The solver retains the attractive feature of being 
positively conservative, typical of the HLL scheme family.

The major improvement over the simple single-state HLL
solver is the ability to resolve contact and tangential 
discontinuities.
This property has been demonstrated by direct comparisons
in several one- and two-dimensional test problems, where
differences are strongly evident.
The results indicate that the new HLLC solver attains sharper
representation of discontinuities, quantitatively similar to the 
exact but algebraically and computationally more intensive 
Riemann solver.

The additional computational cost over the 
traditional HLL approach is less than $8\%$ and we believe
that the improved quality of results largely justifies the trade-off
between the two approximate Riemann solvers.  

Extension to relativistic magnetized flows will be considered 
in a forthcoming paper. We notice, however, that the HLLC
formalism presented in this work can be easily generalized to the 
case of vanishing normal component of magnetic field.
When this degeneracy occurs \cite[as in the propagation of jets with 
toroidal magnetic field, see for example][]{LAAM05}, 
in fact, the solution to the Riemann problem is entirely analogous to 
the non-magnetized case, since only three waves are actually involved.
This extension will be presented in \cite{MMB05}.

Finally, we mention that the relativistic HLLC scheme 
does not make any assumption on the equation of state, 
and efforts to incorporate different equations of state should 
be minimal.

%%%%%%%%%%%%%%%%%%%%%%%%%%%%%%%%%%%%%%%%%%%%%%%%%%%%%%%%%%%%%%%%
%
%          B I B L I O G R A P H Y 
%
%%%%%%%%%%%%%%%%%%%%%%%%%%%%%%%%%%%%%%%%%%%%%%%%%%%%%%%%%%%%%%%%

%%%%%%%%%%%%%%%%%%%%%%%%%%%%%%%%%%%%%%%%%%%%%%%%%%%%%%%%%%%%%%%%%%%%
\appendix
%
%
%%%%%%%%%%%%%%%%%%%%%%%%%%%%%%%%%%%%%%%%%%%%%%%%%%%%%%%%%%%%%%%%%%%%
\section{Proof of $\lambda_L\le\lambda^*\le\Lambda_R$}\label{app:A}

In what follows we prove some important results concerning 
the positivity of the our relativistic HLLC scheme.
The proof is given below in \ref{prop:lambdastar}. Propositions
\ref{prop:AR} through \ref{prop:mixed} demonstrate some preliminary
results. 
   
We assume that the fastest and slowest signal velocities 
are computed according to the prescription given in \S\ref{sec:speeds},
and that $\lambda_R >0$ and $\lambda_L <0$, which is the case
of applicability for the intermediate fluxes (\ref{eq:hllc_flux}).
Obviously, the initial left and right states are supposed
to be physically admissible,
i.e. they belong to the set $G$ defined in \S\ref{sec:positivity}.

%It is assumed that the wave speeds 
%assume that $\lambda_R > 0$ and
%$\lambda_L < 0$.  
%First of all, we notice that the coefficient $A$ appearing 
%in eq. (\ref{eq:lambda_p}) is always positive for $A = A_R$ and
%negative for $A = A_L$. 

%%%%%%%%%%%%%%%%%%%%%%%%%%%%%%%%%%
\begin{proposition}\label{prop:AR}              
 $A_R > 0$, $A_L < 0$            
\end{proposition}                
%%%%%%%%%%%%%%%%%%%%%%%%%%%%%%%%%%

\begin{demonstration} % <<<<<<<<<<<<<<<<<<<<<<<<<<<<<<<<<<

We will only prove $A_R >0$, since the proof for $A_L$
is similar. For the sake of clarity, we omit the subscript 
$R$.
Using the definition of $A$ given after equation (\ref{eq:lambda_p})
one has
\begin{equation}\label{eq:App_A}
  A = (E + p)\left[\lambda\left(1 - \frac{\sigma}{\Gamma}\right) - v_x\right]\,,
\end{equation}
where $\sigma = c_s^2/\gamma^2$ is always positive
numbers. Equation (\ref{eq:App_A}) is always positive for 
$\lambda > \lambda_0$, where  
\begin{equation}\label{eq:App_lambda0}
 \lambda_0 \equiv \frac{\Gamma v_x}{\Gamma - \sigma}\,.
\end{equation} 
However, according to the choice given in \S\ref{sec:speeds},
$\lambda$ must satify 
\begin{equation}\label{eq:App_lambda}
 f(\lambda) = (\lambda - v_x)^2 - \frac{\sigma}{1 - c_s^2}(1 - \lambda^2) 
              \ge 0 \,.
\end{equation}
with $\sigma_s = \sigma/(1-c_s^2)$.
Equation (\ref{eq:App_lambda}) simply states that 
$\lambda$ must be greater than the root with the positive 
sign $\lambda_+$ 
(for the left state, $\lambda_L$ is always less than 
the root with the negative sign $\lambda_-$).
Direct substitution of $\lambda_0$ from (\ref{eq:App_lambda0}) into 
(\ref{eq:App_lambda}) shows, after some algebra, that 
\begin{equation}
 f(\lambda_0) \le K_1\left[-c_s^4 + (v_x^2 + 2\Gamma)c_s^2 - \Gamma^2\right]\,,
\end{equation}
where the equality occurs in the limit of zero tangential velocities
and $K_1$ is always a positive quantity.
Since $1 < \Gamma < 2$ and $c^2_s$ has the limiting
value $(\Gamma - 1)$ the expression in square bracket is always 
negative, which means that $\lambda \ge \lambda_+ > \lambda_0$.
This implies that $A \equiv A_R$ is always positive with our 
choice of $\lambda\equiv\lambda_R$.

Since one can prove, in a similar way, that $A_L < 0$, we
have the important results that $U_E^{hll} = A_R - A_L > 0$.
\end{demonstration}  % <<<<<<<<<<<<<<<<<<<<<<<<<<<<<<<<<<

%%%%%%%%%%%%%%%%%%%%%%%%%%%%%%%%%%%%%%%%
\begin{proposition}\label{prop:AB}
 $B_R + A_R > 0$, $B_L - A_L > 0$.
\end{proposition}
%%%%%%%%%%%%%%%%%%%%%%%%%%%%%%%%%%%%%%%%
\begin{demonstration} % <<<<<<<<<<<<<<<<<<<<<<<<<<<<<<<<<<

Again we only give the proof for the right state, the other case
being similar.
The function $A+B$ (the subscript $R$ is omitted), with $A$ and $B$ defined 
after equation (\ref{eq:lambda_p}) increases linearly with $\lambda$ and 
is positive for $\lambda>\lambda_0$, where
\begin{equation}
 \lambda_0 = \frac{v_x\gamma^2\Gamma(v_x + 1) + c_s^2}
                  {\gamma^2\Gamma(v_x + 1) - c_s^2} \,.
\end{equation}
However, direct substitution of $\lambda_0$ in (\ref{eq:App_lambda})
shows, after extensive manipulations, that
\begin{equation}
  f(\lambda_0) = - K_2\Big[c_s^4 - (1 + 2\Gamma)c_s^2 + \gamma^2\Gamma^2(1 - v_x^2)\Big]
  \,,
\end{equation}
where $K_2$ is always a positive quantity.
It can be easily verified that the function in square brackets 
is always positive if $c_s^2\in[0,\Gamma - 1]$ and $1<\Gamma<2$.
Thus we must have $B_R/A_R > -1$.
\end{demonstration} % <<<<<<<<<<<<<<<<<<<<<<<<<<<<<<<<<<

%%%%%%%%%%%%%%%%%%%%%%%%%%%%%%%%%%%%%%%%
\begin{proposition}\label{prop:mixed}
 $\lambda_LA_R - A_R < 0$, $\lambda_RA_L - B_L  > 0$.
\end{proposition}
%%%%%%%%%%%%%%%%%%%%%%%%%%%%%%%%%%%%%%%%
\begin{demonstration} % <<<<<<<<<<<<<<<<<<<<<<<<<<<<<<<<<<

For the right state we have that 
\begin{equation}\label{eq:App_last}
 \lambda_LA - B \le \lambda_- A - B \le 
 \left(\frac{2v}{1 + \sigma_s} - \lambda\right)A - B \,,
\end{equation}
where the last 
inequality follows from the fact that the
two roots of equation (\ref{eq:App_lambda}) satisfy 
\begin{equation}
 \lambda_- = \frac{2v}{1 + \sigma_s} - \lambda_+ \;, 
  \quad\textrm{and}\quad
 \lambda_+ \le \lambda \,.
\end{equation} 
Using the fact that $\lambda^2(1+\sigma_s) \ge 2\lambda v - v^2 + \sigma_s$
and that $A>0$, the last expression in equation (\ref{eq:App_last}) 
can be shown to obey the following  
\begin{equation}
 \left(\frac{2v}{1 + \sigma_s} - \lambda\right)A - B \le g\,,
\end{equation}
where 
\begin{equation}\label{eq:App_g}
 g = K_3 \Big[v^2(\Gamma - c_s^2 - 1) + 1 - \Gamma + c_s^2 - 2c_s^2v_t^2\Big]\,,
\end{equation}
with $K_3$ being a positive quantity.
The expression in square bracket in equation (\ref{eq:App_g}) 
is always negative under the same assumptions used
previously. Thus we have $\lambda_L < B_R/A_R$ and, similarly,
one can prove that $\lambda_R > B_L/A_L$.

\end{demonstration} % <<<<<<<<<<<<<<<<<<<<<<<<<<<<<<<<<<

%%%%%%%%%%%%%%%%%%%%%%%%%%%%%%%%%%%%%%%%
\begin{proposition}\label{prop:lambdastar}
 $\lambda_L \le \lambda^* \le \lambda_R$.
\end{proposition}
%%%%%%%%%%%%%%%%%%%%%%%%%%%%%%%%%%%%%%%%
\begin{demonstration} % <<<<<<<<<<<<<<<<<<<<<<<<<<<<<<<<<<

We now show that the choice of eigenvalues given 
in \S\ref{sec:speeds} always guarantees 
$\lambda_L \le \lambda^* \le \lambda_R$.

The starting point is to note that the quadratic equation 
(\ref{eq:lambdastar}) can be more conveniently written as
\begin{equation}
  (A_L\lambda^* - B_L)(1 - \lambda_R\lambda^*) =
  (A_R\lambda^* - B_R)(1 - \lambda_L\lambda^*) \,,
\end{equation}
which defines the intersection of two quadratic functions.
The parabola on the left hand side vanishes in $\lambda^* = 1/\lambda_R > 1$
and $\lambda^* = B_L/A_L < 1$, whereas the parabola on the right hand side 
in $\lambda^* = 1/\lambda_L < -1$ and $\lambda^* = B_R/A_R > -1$.
Moreover the two quadratics have the same concavity, since
$\sign(A_L\lambda_R) = \sign(A_R\lambda_L) = -1$.
Thus the intersection must necessarily satisfy
\begin{equation}
 \min\left(\frac{B_R}{A_R}, \frac{B_L}{A_L}\right)\le \lambda^*\le 
 \max\left(\frac{B_R}{A_R}, \frac{B_L}{A_L}\right) \,.
\end{equation}
However, for any $\lambda\in(-1,1)$ one has 
\begin{equation}
  \lambda A - B = (\lambda - v_x)^2(E+p) + p(1 - \lambda^2) > 0  \,,
\end{equation}  
which, together with the results previously shown, implies that  
\begin{equation} \begin{array}{l}
 \DS  1 > \lambda_R > \max\left(\frac{B_R}{A_R},\frac{B_L}{A_L}\right)  \,, \\ \noalign{\medskip}
 \DS  -1 < \lambda_L < \min\left(\frac{B_R}{A_R},\frac{B_L}{A_L}\right) \,. 
 \end{array} 
\end{equation}
and hence
\begin{equation}
  \lambda_L \le \lambda^* \le \lambda_R \,.
\end{equation}

\end{demonstration} % <<<<<<<<<<<<<<<<<<<<<<<<<<<<<<<<<<

\label{lastpage}
\end{document}